\documentclass[runningheads]{llncs}
\usepackage[T1]{fontenc}
\usepackage{graphicx}
\usepackage{makecell}
\usepackage{multirow}
\usepackage{multicol}
\usepackage{booktabs}
\usepackage{amsmath,amssymb,amsfonts}

\usepackage{xcolor}  
\definecolor{myRed}{RGB}{195,10,10}
\definecolor{myGreen}{RGB}{55,149,73}
\definecolor{myBlue}{RGB}{5,5,60}

\begin{document}
\title{
Operating Room Workflow Analysis via Reasoning Segmentation over Digital Twins
}
%
%
\author{Yiqing Shen \and Chenjia Li \and Bohan Liu \and Cheng-Yi Li \and Tito Porras \and Mathias Unberath}
%
%
\institute{Johns Hopkins University, Baltimore, MD, USA\\
\email{yshen92@jhu.edu}}
%
\maketitle              

\begin{abstract}

Analyzing operating room (OR) workflows to derive quantitative insights into OR efficiency is important for hospitals to maximize patient care and financial sustainability. 
Prior work on OR-level workflow analysis has relied on end-to-end deep neural networks. 
While these approaches work well in constrained settings, they are limited to the conditions specified at development time and do not offer the flexibility necessary to accommodate the OR workflow analysis needs of various OR scenarios (\textit{e}.\textit{g}., large academic center \textit{vs.} rural provider) without data collection, annotation, and retraining. 
Reasoning segmentation (RS) based on foundation models offers this flexibility by enabling automated analysis of OR workflows from OR video feeds given only an implicit text query related to the objects of interest. 
Due to the reliance on large language model (LLM) fine-tuning, current RS approaches struggle with reasoning about semantic/spatial relationships and show limited generalization to OR video due to variations in visual characteristics and domain-specific terminology.
To address these limitations, we first propose a novel digital twin (DT) representation that preserves both semantic and spatial relationships between the various OR components.
Then, building on this foundation, we propose ORDiRS (\underline{O}perating \underline{R}oom \underline{Di}gital twin representation for \underline{R}easoning \underline{S}egmentation), an LLM-tuning-free RS framework that reformulates RS into a ``reason-retrieval-synthesize'' paradigm.
Finally, we present ORDiRS-Agent, an LLM-based agent that decomposes OR workflow analysis queries into manageable RS sub-queries and generates responses by combining detailed textual explanations with supporting visual evidence from RS.
Experimental results on both an in-house and a public OR dataset demonstrate that our ORDiRS achieves a cIoU improvement of 6.12\%-9.74\% compared to the existing state-of-the-arts. 

\end{abstract}

\section{Introduction}
Operating room (OR) efficiency is important for optimizing surgical throughput, resource utilization, and patient safety, with direct implications for hospital financial sustainability \cite{friedman2006increasing}.
Video analysis provides objective monitoring of OR workflow patterns and resource utilization, generating data-driven insights that may enable operational improvements \cite{pasquer2024operating}.
However, traditional deep learning methods for OR video analysis are limited to narrowly defined closed-set tasks, which are typically formulated in an end-to-end manner \cite{o2020deep}.
These previous methods can identify surgical instruments or track staff movements, but their closed-set nature 1) limits their flexibility inhibiting their utility for the various clinical settings these algorithms may be used in, and 2) results in them failing to interpret open-set text queries that require reasoning.
To overcome these limitations, reasoning segmentation (RS) \cite{lisa} was introduced. 
Unlike traditional segmentation methods, RS processes text queries that only implicitly refer to the required segmentation targets, thus requiring both spatial and semantic understanding \cite{lisa,visa,llmseg,xia2024gsva}.
Thus, RS may enable automated OR efficiency analysis by identifying workflow bottlenecks, monitoring safety protocol compliance, and evaluating  team coordination during procedural steps \cite{vladu2024enhancing,harari2024deep}

However, current RS approaches, including LISA \cite{lisa,yang2023improved}, VISA \cite{visa}, Reason3D \cite{chen2024reasoning3d}, rely heavily on fine-tuning multimodal large language models (MLLMs) for both perception and reasoning. 
These models struggle with implicit text queries that require multi-step reasoning across semantic and spatial relationships due to inherent limitations in MLLM token representations, which arbitrarily discretize continuous physical information in the OR.
This limitation further hinders OR efficiency analysis, which depends on processing complex sequences of entity interactions while maintaining spatial awareness.
Moreover, these RS models struggle with domain adaptation in OR settings, as they are typically developed and fine-tuned on general-purpose datasets like \textit{ReasonSeg} \cite{lisa}, which differ from medical environments in both visual characteristics and domain-specific terminology \cite{vm2024fine}. 
Even with OR-specific fine-tuning, generalization remains challenging due to data distribution variability across hospitals and healthcare institutions. 
Finally, reliance on MLLM fine-tuning requires frequent updates to maintain compatibility with evolving backbones, potentially disrupting continuous monitoring and increasing implementation costs in OR environments \cite{xia2024understanding}.

To overcome these limitations, we introduce a digital twin (DT) representation that models the OR environment from video.
Our DT representation integrates specialized vision models, such as SAM for identifying and segmenting objects in the OR \cite{sam1,sam2} and DepthAnything2 for estimating spatial depth \cite{depthanything}, to capture and encode semantic and spatial relationships among personnel, equipment, and activities.
We then propose a tuning-free RS approach that restructures RS into a ``reason-retrieve-synthesize'' paradigm, enabling adaptive segmentation without fine-tuning.
The large language model (LLM) performs zero-shot reasoning on the DT representation in three steps: (1) interpreting the implicit text query, (2) analyzing relevant objects and their spatial and semantic relationships, and (3) generating segmentation masks.
For example, when tasked with segmenting non-essential personnel entering sterile fields, the LLM first interprets the query requirements, then identifies relevant staff members by analyzing semantic relationships in the DT representation, evaluates their spatial positions relative to sterile zones, and generates segmentation masks accordingly. 
In contrast to prior approaches to RS that leverage MLLMs to process query and video directly, our approach separates perception from reasoning, allowing the LLM to focus on high-level analysis using structured information from the DT representation.

The major contributions are three-fold.
First, we propose a novel DT representation for OR video that captures and preserves semantic and spatial relationships.
Second, we introduce a tuning-free RS approach for OR data, called Operating Room Digital twin representation for Reasoning Segmentation (ORDiRS), which leverages the DT representation as an intermediate layer to separate perception from reasoning in the LLM.
Third, building on ORDiRS, we propose \textbf{ORDiRS-Agent}, an agent framework designed specifically for OR efficiency analysis.
It first breaks down user queries into manageable RS sub-queries for ORDiRS to process.
It then aggregates the RS results to generate multi-modal responses, combining textual explanations with corresponding RS masks to enhance transparency.

\section{Methods}

\subsubsection{OR Digital Twin Representation Construction}

\begin{figure}[t!]
    \centering
    \includegraphics[width=0.85\linewidth]{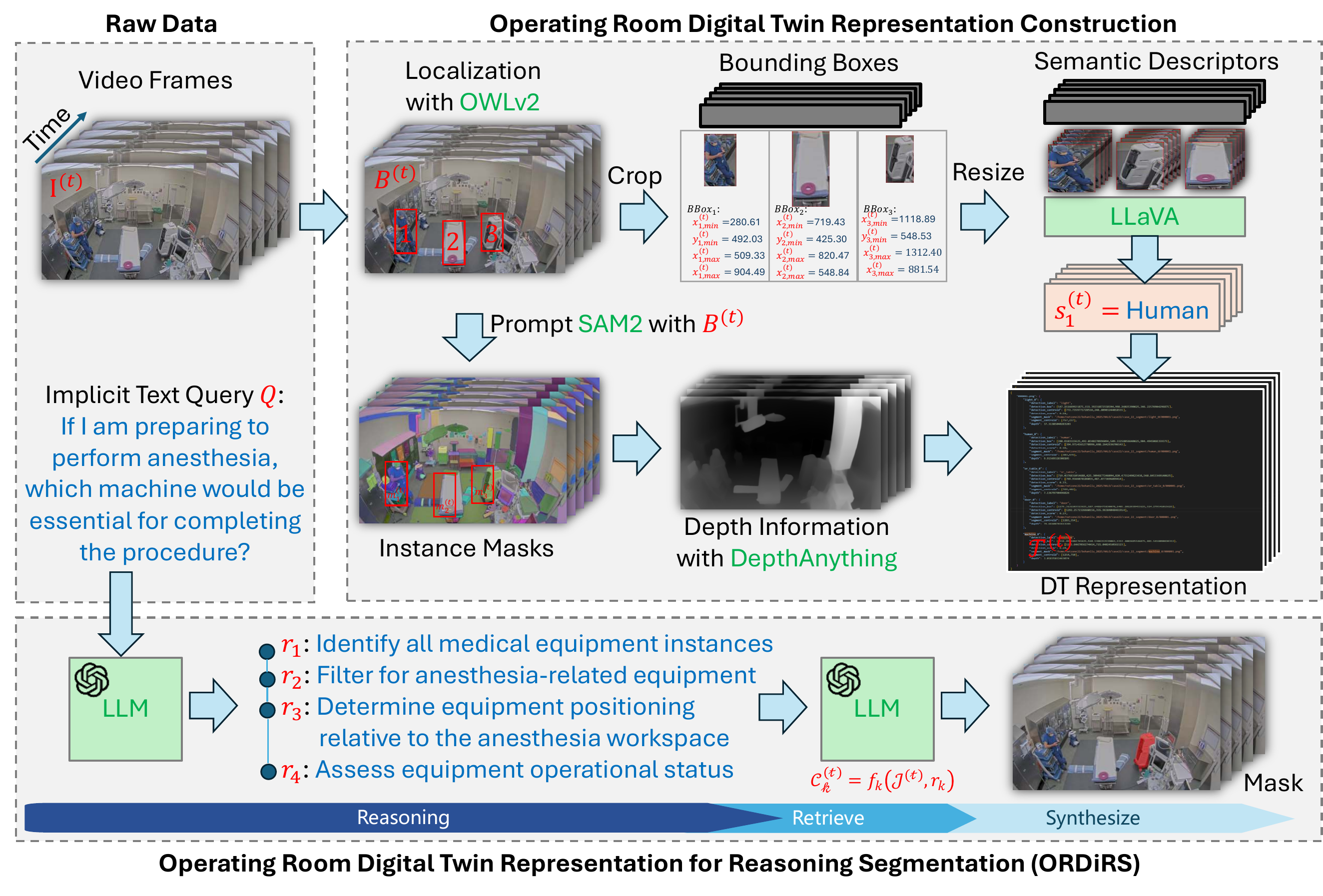}
    \caption{
    Overview of the ORDiRS framework.
    The pipeline consists of two main components.
    (1) DT representation construction: Processing raw OR video frames through multiple vision foundation models, culminating in a structured JSON; 
    (2) RS stage: Implementing a three-stage ``reason-retrieve-synthesize'' paradigm where LLM-based reasoning decomposes an implicit text query into atomic reasoning requirements. 
    %
    }\label{fig:framework}
\end{figure}

Our DT representation creates a representation of the OR environment by integrating outputs from multiple specialized vision foundation models to preserve both semantic relationships between OR components and their spatial configurations, as shown in Fig.~\ref{fig:framework}.
We process the OR video sequence $I = [I^{(1)},I^{(2)},\cdots,I^{(T)}]$ frame by frame, where each frame $I^{(t)}$ at time $t$ is encoded into a structured JSON $\mathcal{J}^{(t)}$. 
We select JSON as our DT representation format due to its hierarchical structure that naturally accommodates nested relationships between frame-level metadata and instance-level attributes while maintaining compatibility with LLMs and enabling efficient querying of spatial-temporal information.

The DT representation construction process begins with OWLv2 \cite{owlv2}, an open vocabulary detection model that generates a set of bounding boxes $B^{(t)} = \{b_i^{(t)}\}_{i=1}^{N}$ for each frame $I^{(t)}$, where each $b_i^{(t)} = [x_{i,\text{min}}^{(t)}, y_{i,\text{min}}^{(t)}, x_{i,\text{max}}^{(t)}, y_{i,\text{max}}^{(t)}]$ represents the spatial coordinates of a detected object with a confidence score $\alpha_i^{(t)}\in[0,1]$. 
%
%
Using each bounding box $b_i^{(t)}$ as a prompt, we employ the SAM2 \cite{sam2} to generate precise instance segmentation masks $M^{(t)} = \{m_i^{(t)}\}_{i=1}^{N}$, where $m_i^{(t)}$ is a binary mask indicating pixel-wise object boundaries with confidence score $\beta_i^{(t)}$ derived from the IoU prediction head in SAM.
%
%
For semantic understanding, we use LLaVA-7B to analyze image regions cropped from $B^{(t)}$ \cite{llava}. 
LLaVA-7B generates semantic descriptors $S^{(t)} = {s_i^{(t)}}_{i=1}^{N}$, where each $s_i^{(t)}$ contains natural language descriptions that capture object attributes, roles, and contextual relationships within the OR environment.
%
%
Analogously, spatial relationships are encoded through depth information using DepthAnything2 \cite{depthanything}, which generates a dense depth map $D^{(t)}$ for each frame uniformly. 
For each instance $i$, we compute depth statistics within its mask region as $d_i^{(t)} = \{D^{(t)}(p) | p \in m_i^{(t)}\}$, where $p$ represents pixel coordinates. 
The mean depth $\mu_i^{(t)}$ and standard deviation $\sigma_i^{(t)}$ are calculated from $d_i^{(t)}$ to characterize the instance's position and depth variation. 
This depth information enables monitoring of spatial relationships for maintaining proper distances between sterile and non-sterile zones in the OR environment.

\subsubsection{Reasoning Segmentation with DT Representation}

Given the $\mathcal{J}^{(t)}$ and an implicit text query $Q$, our RS approach (ORDiRS) follows a three-stage paradigm: ``reason-retrieve-synthesize'' (Fig.~\ref{fig:framework}). 
This design enables zero-shot reasoning by leveraging LLMs' inherent capabilities while maintaining interpretability through structured intermediate outputs.
The reasoning stage begins by decomposing the implicit query $Q$ into explicit reasoning requirements $R = \{r_k\}_{k=1}^K$, where $K$ represents the number of atomic requirements identified by the LLM through chain-of-thought prompting. 
Each requirement $r_k$ specifies the semantic or spatial conditions that the target instances must satisfy. 
%
%
The retrieval stage processes these requirements against the DT representation through a series of filtering operations. 
For each requirement $r_k$, we define a filtering function $f_k$ that evaluates instances based on their attributes in $\mathcal{J}^{(t)}$:
$\mathcal{C}_k^{(t)} = f_k(\mathcal{J}^{(t)}, r_k)$,
where $\mathcal{C}_k^{(t)}$ represents the set of candidate instances satisfying requirement $r_k$ at time $t$. 
To facilitate semantic filtering, $f_k$ leverages the LLM's natural language understanding capabilities by constructing prompts that evaluate instance descriptions $s_i^{(t)}$ against semantic requirements. 
For spatial filtering, $f_k$ prompts LLM to utilize the depth information and instance mask to assess relative positions between instances by generating relative codes.
The final candidate set $\mathcal{C}^{(t)}$ is obtained through the intersection of all requirement-specific candidates \textit{i}.\textit{e}., $\mathcal{C}^{(t)} = \bigcap_{k=1}^K \mathcal{C}_k^{(t)}$.
The synthesis stage generates the final segmentation by combining instance masks from the candidate set. 
For each frame $t$, we construct the reasoning segmentation mask $M_Q^{(t)}$ as $M_Q^{(t)} = \bigcup_{i \in \mathcal{C}^{(t)}} m_i^{(t)}$,
where $m_i^{(t)}$ represents the instance mask for candidate $i$.

\begin{figure}[t!]
    \centering
    \includegraphics[width=0.8\linewidth]{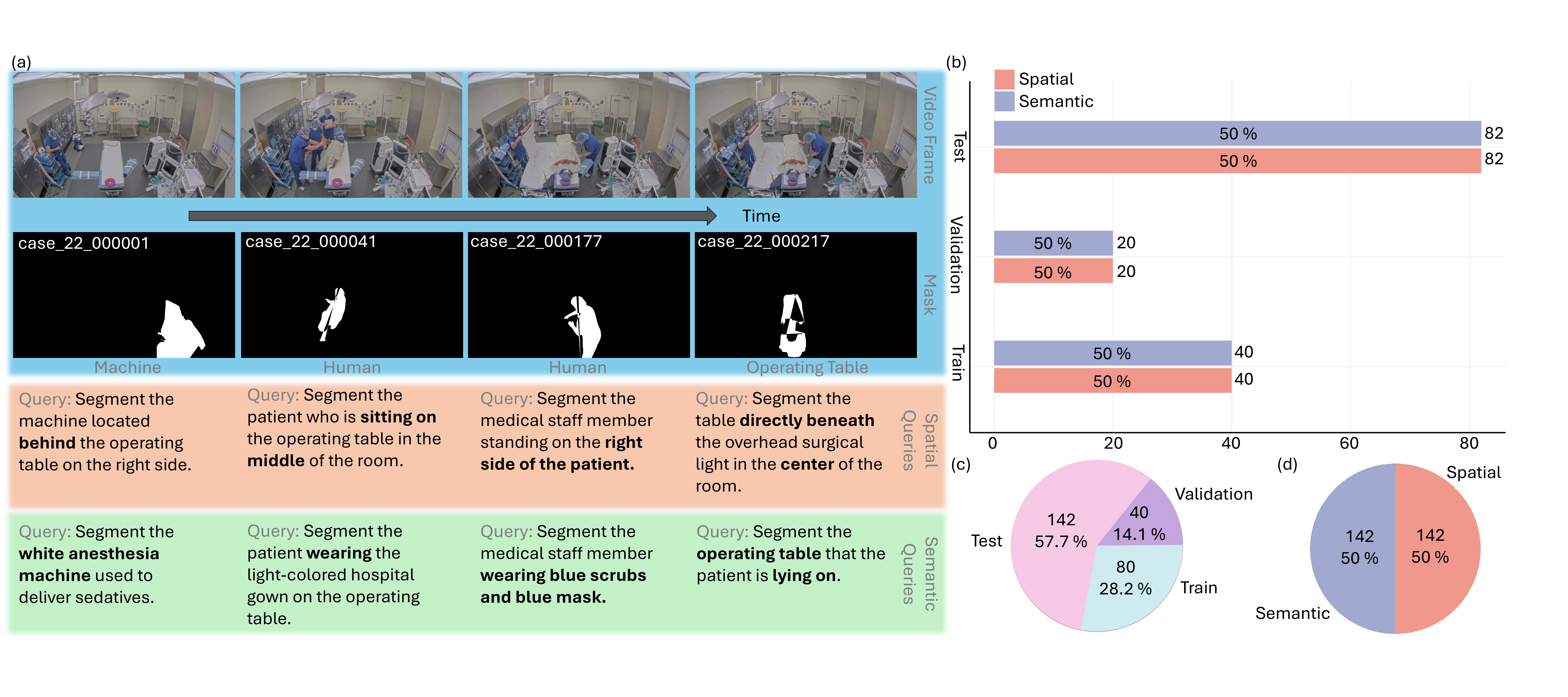}
    \caption{
    Visualization of the OR reasoning segmentation in-house benchmark dataset. 
    (a) Sample video frame sequence from case ID 22 demonstrating paired video frames, segmentation masks, and corresponding spatial/semantic implicit text queries. 
    (b) Query type distribution per dataset split showing balanced representation.
    (c) Dataset split proportions across train, validation, and test datasets. 
    (d) Overall equal distribution between spatial and semantic queries.
    }\label{fig:dataset}
\end{figure}

\subsubsection{Agent for OR Workflow Analysis}

Based on our RS method, we propose ORDiRS-Agent to enable OR workflow analysis through natural language interaction.
Given a user query $Q_\text{OR}$ about OR efficiency, ORDiRS-Agent first decomposes it into a set of RS sub-queries $\{Q_1, Q_2, ..., Q_L\}$ through LLM-based planning.
We implement the LLM-based planning process by a two-step chain-of-thought prompting, where the LLM first analyzes $Q_\text{OR}$ to identify key efficiency aspects that require investigation, then generates targeted RS sub-queries for each aspect. 
%
%
After LLM-based planning, we construct the DT representation $\{\mathcal{J}^{(t)}\}_{t=1}^T$ across the entire OR video sequence and leverage ORDiRS to process each sub-query $Q_l$ to generate corresponding RS masks $\{M_{Q_l}^{(t)}\}_{t=1}^T$. 
%
%
Then, we define an efficiency analysis function $\mathcal{E}_l$ for each sub-query:
$\mathcal{A}_l = \mathcal{E}_l(
\{M_{Q_l}^{(t)}\}_{t=1}^T,
\{\mathcal{J}^{(t)}\}_{t=1}^T)$,
where $\mathcal{A}_l$ represents the analysis results containing temporal patterns, statistical measures, and identified objects. 
The efficiency analysis function $\mathcal{E}_l$ operates in two modes depending on the analysis requirements of $Q_l$. 
For semantic analysis, such as evaluating team coordination patterns or protocol compliance, $\mathcal{E}_l$ directly employs the LLM to reason over the sequence of RS masks and DT representations.
For quantitative analysis tasks, such as computing door opening frequencies or phase transition durations, $\mathcal{E}_l$ utilizes the LLM to generate Python code that processes the mask sequences and extracts relevant statistical metrics. 
%
%
Finally, ORDiRS-Agent aggregates all analysis results through LLM to obtain the final response to $Q_\text{OR}$ by $\mathcal{R} = \texttt{LLM}(\{\mathcal{A}_l\}_{l=1}^L, Q_{OR})$, where $\mathcal{R}$ represents the final response containing both textual explanations and supporting visual evidence in terms of the most relative RS mask. 
%


\section{Experiments}

\begin{table}[!t]
\caption{
The comparison of RS across different reasoning categories (semantic, spatial, mixed) on the in-house dataset. 
The values are reported as mean $\pm$ standard deviation. 
The efficiency metric is the mean inference time per image in seconds.
The upward arrow ($\uparrow$) signifies that higher values indicate better performance.
}
\label{table:expjf}
\centering
\resizebox{\linewidth}{!}{%
\begin{tabular}{l|ccc|ccc|c}
\toprule
\multirow{2}{*}{\textbf{Methods}} 
& \multicolumn{3}{c|}{\textbf{cIoU} ($\uparrow$)}
& \multicolumn{3}{c|}{\textbf{gIoU} ($\uparrow$)} 
& \multirow{2}{*}{\textbf{Time(s)}}  \\
\cline{2-7}
& \textbf{Semantic} & \textbf{Spatial} & \textbf{Mixed}
& \textbf{Semantic} & \textbf{Spatial} & \textbf{Mixed} \\
\hline

LISA-7B-SAM1 \cite{lisa} & 25.17{\tiny$\pm6.34$} & 59.93{\tiny$\pm5.07$} & 39.85{\tiny$\pm$5.34} 
& 34.94{\tiny$\pm$7.96} & 63.77{\tiny$\pm$4.40} & 48.97{\tiny$\pm$4.89} 
& 1.51{\tiny$\pm$0.16} \\

LISA-13B-SAM1 \cite{lisa}  & 28.15{\tiny$\pm4.36$} & 65.35{\tiny$\pm5.66$} & 42.08{\tiny$\pm$3.87} 
& 34.19{\tiny$\pm$4.68} & 68.85{\tiny$\pm$4.77} & 50.75{\tiny$\pm$4.19} 
& 2.03{\tiny$\pm$0.21} \\

LISA-7B-SAM1 (ft) \cite{lisa} & 69.68{\tiny$\pm3.35$} & 69.00{\tiny$\pm$5.53} & 68.30{\tiny$\pm$3.03} 
& 64.62{\tiny$\pm$3.59} & 65.38{\tiny$\pm$5.50} & 64.56{\tiny$\pm$3.06} 
& 1.52{\tiny$\pm$0.11} \\

LISA-13B-SAM1 (ft) \cite{lisa} & 53.32{\tiny$\pm$6.57} & 64.84{\tiny$\pm$5.92} & 59.38{\tiny$\pm$5.09}
& 54.33{\tiny$\pm$6.13} & 65.25{\tiny$\pm$5.86} & 58.19{\tiny$\pm$5.79}
& 1.85{\tiny$\pm$0.10} \\

LISA-7B-SAM2 (ft) \cite{lisa} & 51.14{\tiny$\pm$5.00} & 51.42{\tiny$\pm$4.13} & 52.81{\tiny$\pm$3.87} 
& 43.84{\tiny$\pm$5.94} & 46.99{\tiny$\pm$4.16} & 47.05{\tiny$\pm$3.91} 
& 1.42{\tiny$\pm$0.13} \\

LISA-13B-SAM2 (ft) \cite{lisa} & 49.06{\tiny$\pm$5.87} & 49.45{\tiny$\pm$3.69} & 51.21{\tiny$\pm$4.48} 
& 39.22{\tiny$\pm$6.09} & 42.00{\tiny$\pm$4.23} & 42.69{\tiny$\pm$4.71} 
& 2.09{\tiny$\pm$0.19} \\

V* \cite{wu2024v} & 3.09{\tiny$\pm$1.44} & 3.35{\tiny$\pm$1.25} & 3.26{\tiny$\pm$1.00} 
& 2.93{\tiny$\pm$1.48} & 3.27{\tiny$\pm$1.26} & 3.15{\tiny$\pm$1.06} 
& 22.62{\tiny$\pm$0.39} \\

\hline
\textbf{Ours} & \textbf{75.80{\tiny$\pm$3.58}} & \textbf{78.74{\tiny$\pm$3.19}} & \textbf{77.25{\tiny$\pm$2.34}} 
& \textbf{76.85{\tiny$\pm$3.23}} & \textbf{82.48{\tiny$\pm$2.49}} & \textbf{79.55{\tiny$\pm$1.97}}
& 89.05{\tiny$\pm$2.51} \\
\bottomrule
\end{tabular}%
}
\end{table}

\subsubsection{Implementation Details}
%
For LLMs used in our method, we utilize the GPT-4o.
%
%
We leverage the cumulative intersection over the cumulative union (cIoU) and the average of all per-image intersection-over-unions (gIoU) as the metrics for RS \cite{lisa}.
We compare our method against LISA \cite{lisa} (7B and 13B parameter versions) including both the original and fine-tuned LISA (on our dataset) \cite{lisa} with their 7B and 13B parameter variants; and V* \cite{wu2024v}. 
Each LISA variant was evaluated using two different SAMs, \textit{i}.\textit{e}., SAM1 \cite{sam1} and SAM2 \cite{sam2}.
%

\subsubsection{Benchmark Dataset}
To evaluate RS performance on OR videos, we construct a benchmark dataset from in-house OR video recordings, collected under IRB approval with appropriate consent from all participants. 
The dataset encompasses recordings from four operating rooms (case IDs 22, 25, 26, and 28), specifically focusing on OR workflows centering on patients, anesthesiologists, anesthesia machines, and operating tables.
Two experts manually annotated each frame with ground truth segmentation masks and the corresponding video-level implicit text query, as shown in Fig.~\ref{fig:dataset}(a). 
The annotation process incorporated both spatial and semantic queries, where spatial queries address location-based requirements and semantic queries focus on attribute-based or function-based identification. 
The dataset comprises 142 frames in total, divided into training (1 OR, 40 frames, 28.2\%), validation (1 OR, 20 frames, 14.1\%), and test sets (2 OR, 82 frames, 57.7\%) on OR-level, as illustrated in Fig.~\ref{fig:dataset}(b,c). 
The training and validation sets are only used for the baseline methods \textit{i}.\textit{e}., LISA \cite{lisa}, and we conduct comparisons of all methods on the test set.
The query composition maintains an equal balance between spatial and semantic types across all splits (Fig.~\ref{fig:dataset}(d)), with each category representing 50\% of the total queries.
In addition to the in-house OR dataset, we extended a publicly available \textit{MOVR} dataset \cite{srivastav2018mvor} of 732 frames with the RS annotations and obtained \textit{MOVR-Reason}, following the same annotation workflow.

\begin{figure}[t!]
    \centering
    \includegraphics[width=\linewidth]{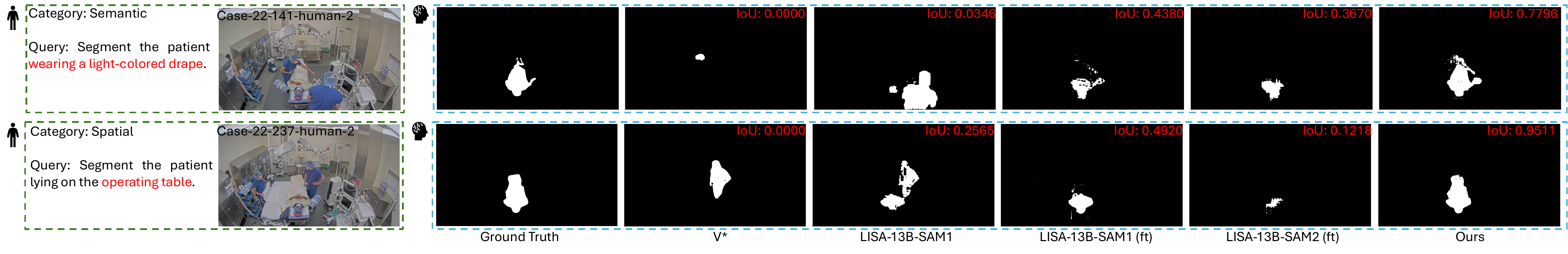}
    \caption{
Qualitative comparison of RS results. 
Two representative cases are shown: a semantic RS task (top row) requiring identification of a patient with specific clothing attributes, and a spatial RS task (bottom row) involving positional understanding. 
%
%
White regions represent the segmentation masks, with ground truth shown in the leftmost column.
%
    }\label{fig:result}
\end{figure}

\subsubsection{Reasoning Segmentation Performance} 
Table \ref{table:expjf} presents the comparison of RS results and the computational efficiency. 
Our method demonstrates superior performance across all reasoning categories. 
Specifically, ORDiRS attains the highest cIoU scores of 75.80\%, 78.74\%, and 77.25\% for semantic, spatial, and mixed reasoning tasks respectively, surpassing the previous best results from LISA-7B-SAM1 (ft) by margins of 6.12\%, 9.74\%, and 8.95\%. 
Similar improvements are observed in gIoU metrics, where our method achieves 76.85\%, 82.48\%, and 79.55\% across the three categories, representing substantial gains over existing approaches. 
However, it comes at the cost of increased computational overhead, with our method requiring 89.05 seconds per image compared to 1.51-22.62 seconds for baseline methods. 
This suggests that our method, without additional improvements to its real-time inference capabilities, is most suitable for offline workflow analysis tasks. 
Similar results can be observed from the \textit{MOVR-Reason} dataset, in Table \ref{table:exp-public}. 
Notably, while fine-tuned LISA variants (denoted by 'ft') show improved performance over their base counterparts, they still fall short of our tuning-free approach, highlighting the effectiveness of our ``reason-retrieval-synthesize paradigm'' in leveraging LLM's inherent reasoning capabilities for OR environment understanding, as shown in Fig.~\ref{fig:result}.

\begin{table}[!t]
\caption{
The comparison of RS across different reasoning categories (semantic, spatial, mixed) on the \textit{MVOR-Reason} dataset.
}
\label{table:exp-public}
\centering
\resizebox{\linewidth}{!}{%
\begin{tabular}{l|ccc|ccc|c}
\toprule
\multirow{2}{*}{\textbf{Methods}} 
& \multicolumn{3}{c|}{\textbf{cIoU} ($\uparrow$)}
& \multicolumn{3}{c|}{\textbf{gIoU} ($\uparrow$)} 
& \multirow{2}{*}{\textbf{Time(s)}}  \\
\cline{2-7}
& \textbf{Semantic} & \textbf{Spatial} & \textbf{Mixed}
& \textbf{Semantic} & \textbf{Spatial} & \textbf{Mixed} \\
\hline

LISA-7B-SAM1 \cite{lisa} & 32.45{\tiny$\pm3.20$} & 24.18{\tiny$\pm6.70$} & 28.12{\tiny$\pm3.20$} 
& 27.51{\tiny$\pm3.04$} & 22.45{\tiny$\pm2.15$} & 24.24{\tiny$\pm1.08$} 
& 1.72{\tiny$\pm0.22$} \\

LISA-13B-SAM1 \cite{lisa}  & 34.25{\tiny$\pm6.20$} & 40.45{\tiny$\pm2.13$} & 38.20{\tiny$\pm1.99$} 
& 32.20{\tiny$\pm3.87$} & 30.05{\tiny$\pm3.43$} & 31.22{\tiny$\pm2.98$} 
& 1.93{\tiny$\pm0.54$} \\

LISA-7B-SAM1 (ft) \cite{lisa} & 45.34{\tiny$\pm2.15$} & 49.45{\tiny$\pm4.56$} & 46.67{\tiny$\pm5.14$} 
& 50.20{\tiny$\pm4.74$} & 47.30{\tiny$\pm3.40$} & 49.49{\tiny$\pm5.14$} 
& 1.77{\tiny$\pm0.41$} \\

LISA-13B-SAM1 (ft) \cite{lisa} & 52.45{\tiny$\pm3.54$} & 48.99{\tiny$\pm2.38$} & 51.74{\tiny$\pm1.45$} 
& 56.84{\tiny$\pm3.16$} & 55.39{\tiny$\pm2.15$} & 55.96{\tiny$\pm3.45$} 
& 1.98{\tiny$\pm0.28$} \\

LISA-7B-SAM2 (ft) \cite{lisa} & 56.38{\tiny$\pm$7.21} & 58.20{\tiny$\pm$3.98} & 57.56{\tiny$\pm$4.45} 
& 50.45{\tiny$\pm$3.24} & 56.18{\tiny$\pm$3.06} & 53.49{\tiny$\pm$1.81} 
& 1.98{\tiny$\pm$0.22} \\

LISA-13B-SAM2 (ft) \cite{lisa} & 59.47{\tiny$\pm$2.12} & 61.31{\tiny$\pm$2.54} & 60.26{\tiny$\pm$3.88} 
& 59.32{\tiny$\pm$2.11} & 53.22{\tiny$\pm$2.58} & 58.90{\tiny$\pm$3.68} 
& 2.39{\tiny$\pm$0.42} \\

V* \cite{wu2024v} & 41.22{\tiny$\pm$3.56} & 44.52{\tiny$\pm$3.52} & 43.30{\tiny$\pm$4.98} 
& 39.54{\tiny$\pm$3.45} & 45.76{\tiny$\pm$5.81} & 41.20{\tiny$\pm$4.12} 
& 25.30{\tiny$\pm$2.41} \\

\hline
\textbf{Ours} & \textbf{83.21{\tiny$\pm$2.81}} & \textbf{86.11{\tiny$\pm$3.76}} & \textbf{84.35{\tiny$\pm$3.40}} 
& \textbf{82.54{\tiny$\pm$2.54}} & \textbf{80.19{\tiny$\pm$3.22}} & \textbf{81.54{\tiny$\pm$4.50}}
& 58.21{\tiny$\pm$7.31} \\
\bottomrule
\end{tabular}%
}
\end{table}

\subsubsection{Case Study on Operating Room Efficiency Analysis} 

\begin{figure}[t!]
    \centering
    \includegraphics[width=0.95\linewidth]{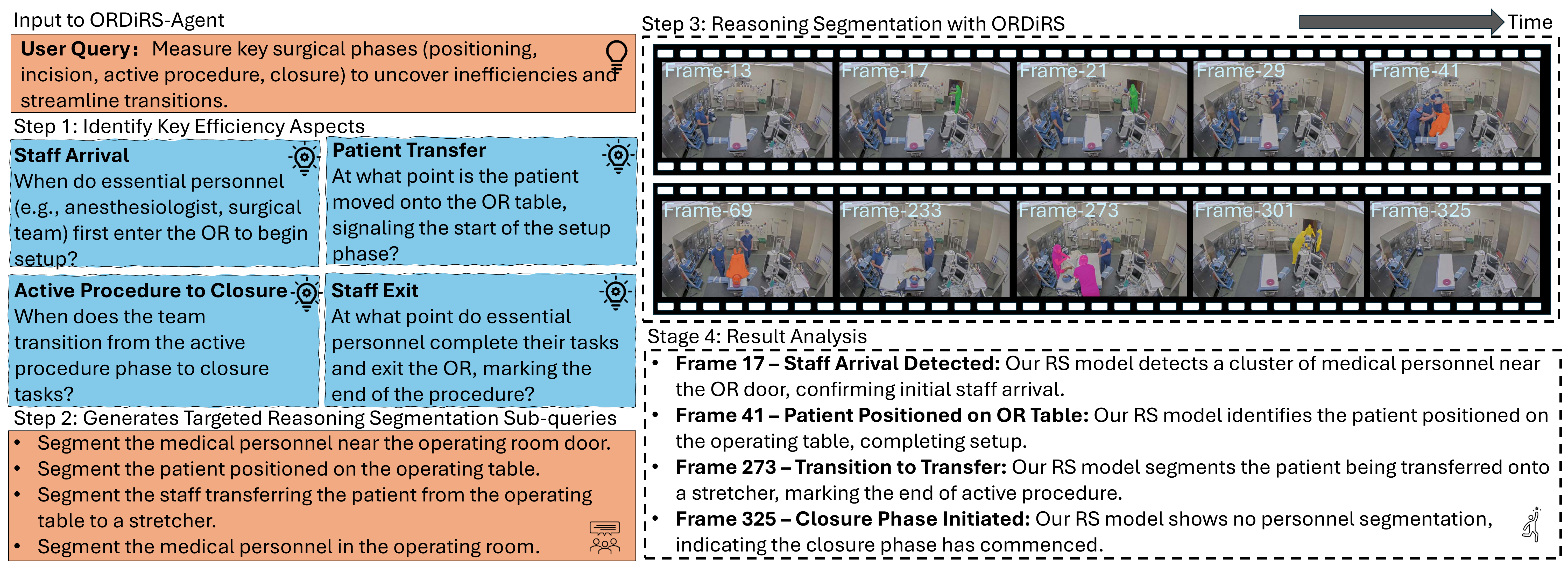}
    \caption{
    A case study for the workflow of ORDiRS-Agent for analyzing operating room efficiency. 
    The process begins with a user query about surgical phase transitions, followed by the identification of key efficiency aspects. 
    It then generates targeted reasoning segmentation sub-queries (Step 2), performs reasoning segmentation using ORDiRS (Step 3), and concludes with result analysis (Step 4). 
    The visualization demonstrates how ORDiRS-Agent tracks critical workflow events across frames, including staff arrival (Frame 17), patient positioning (Frame 41), transition to transfer (Frame 273), and closure phase initiation (Frame 325).
    }\label{fig:case}
\end{figure}

We evaluated ORDiRS-Agent through a case study analyzing surgical workflow efficiency in Fig.~\ref{fig:case}. 
Given the query `\textit{`Measure key surgical phases to uncover inefficiencies and streamline transitions},'' ORDiRS-Agent can detect important transition points. 
It demonstrates ORDiRS-Agent's capability to decompose complex efficiency queries into manageable RS tasks, and provide comprehensive visual-temporal insights.
\section{Conclusion}

We present ORDiRS, an RS framework that utilizes DT representations to analyze OR videos.
By separating perception from reasoning through the DT representation,  ORDiRS eliminates the need for fine-tuning while ensuring robust performance across diverse OR environments. 
We also introduce ORDiRS-Agent, an analytical framework designed to evaluate OR efficiency using ORDiRS.
Our results demonstrate that structured intermediate representations effectively bridge the gap between raw visual data (\textit{e}.\textit{g}., video) and high-level reasoning.
Future work could extend the DT representation framework to capture temporal dependencies and long-term patterns in surgical workflows, enabling deeper efficiency analyses. 
Moreover, the ‘reason-retrieve-synthesize’ paradigm could be adapted for other healthcare settings requiring complex visual reasoning, such as emergency departments and intensive care units.


\bibliographystyle{plain}
\bibliography{6_ref}

\end{document}